\documentclass[prl,twocolumn,showpacs,preprintnumbers,amsmath,amssymb,superscriptaddress]{revtex4}

\usepackage{graphicx}
\usepackage{epsfig}
\usepackage{dcolumn}
\usepackage{bm}

\newcommand{\cng}{CeNiGe$_2$}
\newcommand{\tk}{$T_K$}

\newcommand{\tn}{$T_N$}
\newcommand{\rz}{$\rho_0$}

\begin{document}

\preprint{}

\title{The unusual phase diagram of CeNiGe$_2$}

\author{A.~T.~Holmes}
\email{ath1@st-andrews.ac.uk}
\affiliation{Center for Quantum Science and Technology under Extreme Conditions, Osaka University, Toyonaka, Osaka 560-8531, Japan}
\altaffiliation[Current address: ]{School of Physics and Astronomy, University of St. Andrews, North Haugh, St. Andrews, Fife KY16 9SS, United Kingdom}
\author{T.~Muramatsu}
\affiliation{Center for Quantum Science and Technology under Extreme Conditions, Osaka University, Toyonaka, Osaka 560-8531, Japan}
\author{D.~Kaczorowski}
\affiliation{Institute of Low Temperature and Structure Research, Polish Academy of Sciences, P.O. Box 1410, 50-950 Wroc{\l}aw, Poland}
\affiliation{Center for Quantum Science and Technology under Extreme Conditions, Osaka University, Toyonaka, Osaka 560-8531, Japan}
\author{Z.~Bukowski}
\affiliation{Institute of Low Temperature and Structure Research, Polish Academy of Sciences, P.O. Box 1410, 50-950 Wroc{\l}aw, Poland}
\author{T.~Kagayama}
\affiliation{Center for Quantum Science and Technology under Extreme Conditions, Osaka University, Toyonaka, Osaka 560-8531, Japan}
\author{K.~Shimizu}
\affiliation{Center for Quantum Science and Technology under Extreme Conditions, Osaka University, Toyonaka, Osaka 560-8531, Japan}

\date{\today}

\begin{abstract}
The heavy fermion antiferromagnet CeNiGe$_2$ was investigated under pressure by resistivity and ac calorimetry up to 4$\:$GPa and down to 40$\:$mK.  The two magnetic transitions found in both resistivity and specific heat at 0.1$\:$GPa at $T_{N1}=3.95$ and $T_{N2}=3.21\:$K are replaced by a single one at 0.7$\:$GPa and 2.81$\:$K.  Increasing pressure initially reduces this further, however at about 1.7$\:$GPa a new transition appears, accompanied by a marked change in the pressure dependence of the ordering temperatures, the temperature dependence of the resistivity, and the residual resistivity. There are signs that this new transition has some first order character. The phase diagram of CeNiGe$_2$ bears little resemblance to the Doniach phase diagram widely used to classify heavy fermion compounds.
\end{abstract}

\maketitle


Cerium-based heavy fermion intermetallic compounds have been studied extensively for many years due to their diverse and fascinating range of ground states. The ternary Ce-Ni-Ge family displays the entire range of behavior seen in these systems:  Ferromagnetic, antiferromagnetic (AFM), heavy fermion (HF), superconducting, non-Fermi liquid, and intermediate valent (IV) types of behavior are all found either under ambient conditions or with the application of hydrostatic pressure \cite{Durivault03}. The ground state properties are largely determined by the configuration of the Ce 4f electron: its degree of localization on the Ce ion, and the extent to which its magnetic moment is screened by conduction electrons (the Kondo effect).

A consensus has emerged that the magnetic phase diagram of most individual Ce compounds can be fitted into a single scheme, known as the Doniach diagram, governed by the magnitude of the exchange interaction between the Ce 4f and conduction electrons. This is strongly affected by changes in unit cell volume, which can be controlled by pressure or chemical substitution. Compression tends to move the ground state towards a non-magnetic limit in the sequence AFM-HF-IV.  The theoretical treatment by Doniach \cite{Doniach77} considered the competing RKKY and Kondo interactions, which respectively promote and suppress magnetic order. This model predicts an AFM ordering temperature \tn\ which first increases with pressure, and then is suppressed to zero in a second order manner as the local moments are screened completely.  The point at which \tn\ reaches zero is known as the quantum critical point (QCP). Many novel phenomena have been found at the QCP, such as unconventional superconductivity and non-Fermi liquid behavior.  The heavy fermion antiferromagnet \cng, however, is an exception to this scheme.

We present resistivity and specific heat measurements under high pressure on \cng. This compound lies on the boundary between AFM, HF, and IV behavior in the Ce-Ni-Ge ternary phase diagram. It is a highly anisotropic compound, crystallizing in the orthorhombic CeNiSi$_2$-type structure (space group \emph{Cmcm}).  The easy magnetic direction is parallel to the crystallographic $a$-axis \cite{Pikul04}. The Sommerfeld coefficient $\gamma$ has been estimated to be 100 mJ/mol K$^2$ at $p=0$, though its determination is complicated by the presence of magnetic order \cite{Pecharsky91,Lu05}; at ambient pressure \cng\ shows two antiferromagnetic ordering temperatures, at $T_{N1}=3.2\:$K and $T_{N2}=3.9\:$K.

The isostructural sister compound CeNiSi$_2$ is an intermediate valence system with a unit cell volume about 4\% smaller than \cng.  Several investigations have been carried out on the intermediate alloys CeNi(Ge$_{1-x}$Si$_x$)$_2$, \cite{Hong03a,Hong03b,Kim04,Lu05,Im07}.  These show a disappearance of magnetism for $x\simeq1$, and some signs of a QCP at this point.  They also show a non-linear decrease of $T_N$ with $x$, and a transition to IV behavior as $x\rightarrow1$. By applying pressure to \cng, we wished to reproduce the effects of Si substitution, by reducing the cell volume and suppressing magnetic order, if possible to pass through a QCP and eventually reaching the IV state at high enough pressure.  However, several intriguing phenomena appeared even before these features were reached, and these are subject of this paper.

The sample was selected from single crystals of CeNiGe$_2$ grown by the In-flux method as described previously \cite{Pikul04}. It was cut and polished to $90\times200\times20\:\mu$m$^3$. Six 10$\:\mu$m wires were spot-welded to the sample so that the electrical current for resistivity measurements was oriented along the $a$ axis. They included two \underline{Au}Fe(0.07\%) wires, forming thermocouples at either end of the sample, and a pair of voltage contacts suitable for four-point resistance measurements. Knowledge of the sample geometry enabled us to estimate the absolute resistivity to within $\sim16\%$.

It is generally very difficult to measure specific heat under pressure due to the tiny size of the samples and overwhelming addenda contribution. However, the ac calorimetry method enables the sample specific heat to be determined in a semi-quantitative way, despite these drawbacks \cite{Sullivan68}. An alternating heating current of up to 4.7$\:$mA was applied to one thermocouple, and the resulting temperature oscillations measured at the other via lock-in detection. The system can be modeled as a heat capacity $C$ connected via a thermal resistance $K$ to a bath at temperature $T_0$, giving a characteristic sample relaxation frequency $\omega_{c1}=K/C$. The equilibration time between the thermocouple, heater and sample is combined into a second (higher) characteristic frequency $\omega_{c2}$. At various temperatures the frequency dependence of the thermocouple signal $\tilde V_{ac}$ was measured, and the amplitude fitted to the formula:
\begin{equation}\label{Tac}
    |\tilde V_{ac}|=\frac{A}{\omega} \left(1+(\omega/\omega_{c1})^{-2}+(\omega/\omega_{c2})^2\right)^{-1/2}.
\end{equation}

Provided that the working frequency $\omega\ll\omega_{c2}$, the model can be simplified, neglecting $\omega_{c2}$ and allowing an estimate of the heat capacity $C$ to be extracted from the amplitude and phase of the temperature oscillations \cite{Holmes04a}.  The parameters $\omega_{c1}$ and $\omega_{c2}$ decreased by a factor of 10 from 4.2$\:$K to 1.5$\:$K, and as $\omega_{c2}\sim 2\omega_{c1}$, we chose a working frequency slightly below $\omega_{c1}$ at 1$\:$K, typically $\sim100\:$Hz.  Usually the temperature dependence was measured at two or more frequencies, and the calculated specific heat compared.  A large disagreement indicated a decoupling of the sample and thermocouple, and with a measurement at $\omega\ll\omega_{c1}$ the temperature offset of sample above the background temperature could be estimated, and was typically in the range 0.2--10$\:$mK.

This method has very high sensitivity, but it is difficult to ascertain the absolute value of $C_p$, owing to uncertainties in the thermocouple calibration under pressure, the contribution of the pressure medium, diamonds and wires, and the absolute power delivered to the sample.  However, all of these are likely to vary slowly with pressure, so we can make definite observations of phase transitions, including the shape and size of any jump in specific heat.

High pressure was generated using the clamped diamond anvil cell (DAC) method, with NaCl as a pressure medium. Force was applied at room temperature, and the ruby fluorescence method used to determine the pressure, $p$, at around 25K; we estimated the pressure gradients to be about 10\%. A considerable loss of pressure occurred on cooling from room temperature to around 40$\:$K, however on further cooling little hysteresis was seen in the resistivity, so we believe the pressure remained constant below this temperature. Resistance measurements were carried out in a dilution cryostat, and ac calorimetry in an $^4$He dewar up to 1.4$\:$GPa, and in a dilution cryostat above this pressure. For the latter, the heating current was reduced as the temperature decreased, and the signal scaled appropriately.

In the remainder of this paper, we will first describe the general behavior of \cng\ at effectively ambient pressure, then explain how the ordering temperatures vary with pressure, followed by a detailed look at how $p$ affects the temperature dependence of resistivity and specific heat below 4$\:$K.

\begin{figure}
  \includegraphics[width=\columnwidth]{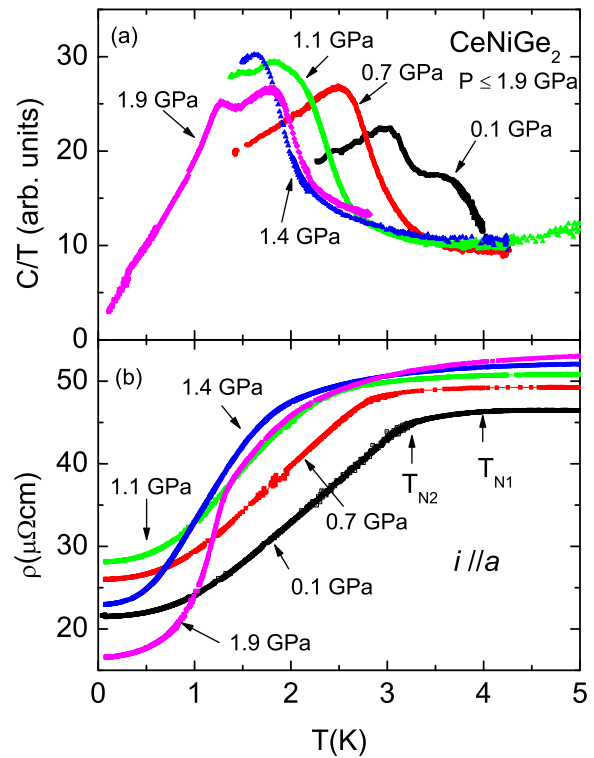}\\
  \caption{(Color online) AC specific heat (a) and resistivity (b) up to 1.9$\:$GPa. The position of the vertical arrows showing $T_{N1}$ and $T_{N2}$ were determined from the inflection points of the specific heat. Note the early disappearance of the second transition, and the dramatic change of behavior in the resistivity between 1.4 and 1.9$\:$GPa.}\label{regionI}
\end{figure}

\begin{figure}
  \includegraphics[width=\columnwidth]{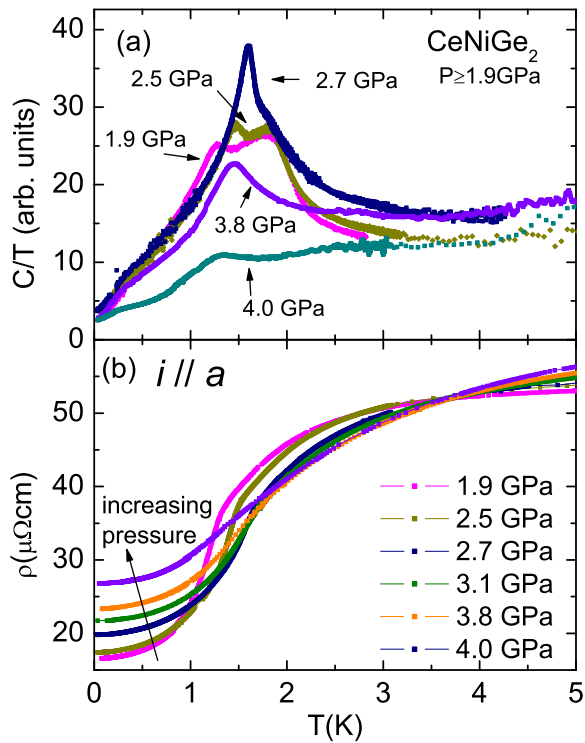}\\
  \caption{(Color online) AC specific heat (a) and resistivity (b) above 1.9$\:$GPa.  Note that the curves at 1.9$\:$GPa are included in both this and fig.~\ref{regionI} for comparison, and the $C_p$ results at 3.1$\:$GPa are omitted from (a) due to experimental difficulties.  }\label{regionII}
\end{figure}

Figures \ref{regionI} and \ref{regionII} show the ac specific heat and resistivity of \cng\ as a function of temperature and pressure, below 5$\:$K.  The compound displays rather different behavior above and below about 1.7$\:$GPa, so the results have been divided for clarity, Fig.~\ref{regionI} showing those up to 1.9$\:$GPa (region I) and Fig.~\ref{regionII} those in the high pressure region II.  The curves at 1.9$\:$GPa are repeated in both figures for comparison.

The resistivity of \cng\ initially decreases on cooling from room temperature, most likely due to a reduction of phonon scattering. This is followed by two broad maxima in $\rho(T)$, at $T_2^{\rm max}$ ($\sim60\:$K) and $T_1^{\rm max}$ ($\sim5\:$K at ambient pressure). This behavior is typical of a Kondo lattice system subject to crystal field splitting of the f-level. Below $T_1^{\rm max}$, magnetic ordering can be discerned, sometimes by a clear kink in the resistivity, or at least by an anomaly observable in $d\rho/dT$.

The two antiferromagnetic transitions at close to ambient pressure are easily identified as separate peaks in the ac specific heat (fig.~\ref{regionI}(a)) at temperatures $T_{N1}=3.95$ and $T_{N2}=3.21\:$K, in agreement with previous reports.  The low temperature kinks in resistivity corresponded exactly to peaks in the specific heat at all pressures up to 4$\:$GPa (or more precisely to the inflection point of the high temperature side of the maximum in $C_p/T$). At 0.1$\:$GPa the anomaly in resistivity at $T_{N1}$ is very weak, and can only be seen clearly by taking the derivative $d \rho / dT$.  There is also broad Kondo maximum $T_1^{\rm max}$ at 4.5$\:$K.

\begin{figure}
  \includegraphics[width=\columnwidth]{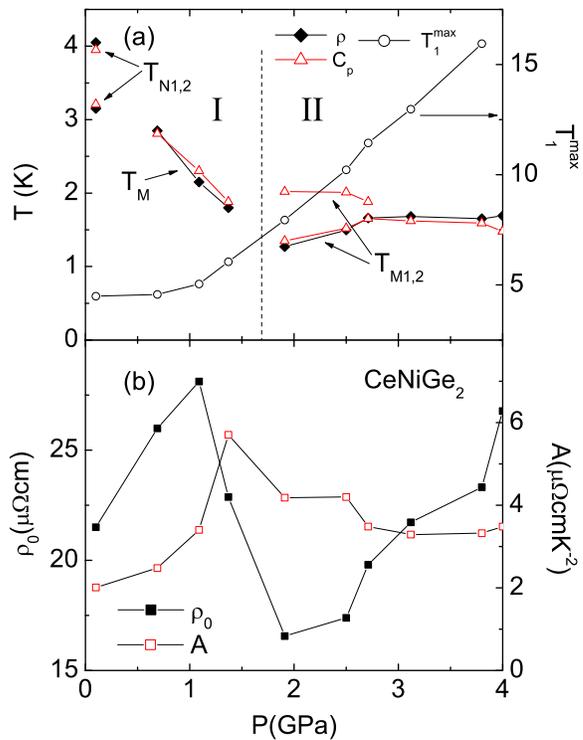}\\
  \caption{(Color online) (a) Left hand scale: Ordering temperature determined by resistivity (filled diamonds) and specific heat (open triangles). Labels as described in text. Right hand scale: Kondo coherence temperature $T_1^{\rm max}$. (b) Coefficients of a fit to $\rho(T)=\rho_0+AT^2$ below 0.5$\:$K.}\label{TNTmaxArho0}
\end{figure}

Pressure affects the transition temperatures as follows (see fig. \ref{TNTmaxArho0}(a)): As the pressure starts to increase, the two transitions are replaced only one; at 0.7$\:$GPa, a single peak can be identified in $C_p$ at a temperature $T_M=2.81\:$K.  On further increasing the pressure, initially $T_M$ decreases linearly with $p$.  At 1.9$\:$GPa a new transition appears at $T_{M2}=1.35\:$K, this is clearly visible in both resistivity and specific heat. A second transition is found at $T_{M1}=2.02\:$K, barely visible in resistivity but with a clear signature in $C_p$. From this pressure onwards the temperatures $T_{M2}$ and $T_{M1}$ no longer decrease monotonically.  $T_{M2}$ rises slightly to a broad maximum of around 1.65$\:$K before saturating at $\sim1.5\:$K, while $T_{M1}$ appears to rise slightly and then converge with $T_{M2}$, though the signature of this transition becomes weaker and is no longer visible above 2.7$\:$GPa.

The low temperature maximum in resistivity $T^{\rm max}_1$ is more or less proportional to the Kondo temperature \tk.  $T^{\rm max}_1$ initially increases quite slowly with pressure, then much faster above $\simeq1\:$GPa.

The temperature dependence of the resistivity as $T\rightarrow0$ gives an indication of the nature of elementary electronic excitations. Below 0.5$\:$K, $\rho(T)$ can be fitted to a power law, $\rho=\rho_0+AT^n$, with $n$ slightly greater than 2. This is typical of Fermi-liquid behavior in the presence of some spin wave excitations. \rz\ reflects scattering from static disorder, and $A$ is determined by dynamic scattering of the quasiparticles, where the bare electron-electron interactions are strongly renormalized by low energy excitations in these systems.    As the pressure is increased up to 1.4$\:$GPa, $A$ rises by nearly a factor of 3 (fig. \ref{TNTmaxArho0}(b)), indicating a increase of the Sommerfeld coefficient from 448 to 755$\:$mJ mol$^{-1}$K$^{-2}$, derived by the Kadowaki-Woods relation for correlated systems \cite{Kadowaki86}.  After the appearance of $T_{M2}$, $A$ decreases markedly, and remains roughly constant up to 4$\:$GPa.

The residual resistivity \rz\ is usually thought to reflect the impurity concentration in a sample, which is not affected by pressure.  The large variations of \rz\ with $p$ in \cng\ are therefore surprising. \rz\ initially starts to increase, however at 1.4$\:$GPa it suddenly falls, continuing to drop further up to 1.9$\:$GPa.  This appears to be related to the appearance of $T_{M2}$. Above this pressure the residual resistivity increases monotonically. This variation of \rz\ is contrasted by the smooth pressure evolution of the resistivity at higher temperature.  It is especially striking to compare the resistivity curves at 1.1 and 1.9$\:$GPa. Above $T_{M2}$ the curves are nearly identical, but it seems like a large bite has been taken out of the resistivity below the transition. Interestingly, the resistivity curve at 1.4$\:$GPa is an intermediate case, with no visible transition, but a drop in \rz, perhaps indicating some sort of precursor effect.

An enhancement of the residual resistivity, and particularly a peak of the $A$ coefficient of the resistivity are often found at an antiferromagnetic QCP, where abundant low energy excitations can strongly renormalize the electronic effective mass, and alter impurity potentials. The low pressure (region I) behavior  of CeNiGe$_2$ strongly resembles the approach to such a QCP, however anomalies associated with magnetic order persist up to the highest pressures so far measured.  Extrapolating the decrease of the ordering temperature from $p=0$ leads to a projected QCP at around 3$\:$GPa.  The system effectively avoids this QCP, with a new transition $T_{M2}$ enabling the large entropy that would otherwise evolve at very low temperature to be absorbed into some new ordered phase.

The pressure evolution of $T_{M2}$ is also rather unexpected.  The transition temperature appears to saturate with increasing pressure, and the specific heat signature first sharpens before growing weaker (see fig.~\ref{regionII}(b)).  The sharpening of the peak implies that pressure gradients are not a serious problem, though to be sure measurements in a hydrostatic medium are necessary. The collapse of the specific heat jump at higher pressure could be explained by phase separation, if the transition at $T_{M2}$ has some first order character.

The drop in residual resistivity above 1.4$\:$GPa, accompanied by the appearance of a more sharply defined kink in resistivity seems at first to be slightly paradoxical.  A more fully gapped Fermi surface might produce a stronger resistivity signature, but it should also increase, rather than decrease \rz\ due to the decrease in density of states.  In \cng\ we see the opposite. However, the effect might be explained by a reorientation of the magnetic order, causing the scattering along the observed current direction to become more effective.

The smooth variation of the resistivity with pressure at high temperature, up to 300$\:$K, with no change in the overall shape, implies that there were no problems with the electrical contacts, which could otherwise cause shifts in the resistance measured and hence the inferred absolute resistivity.  It also rules out structural changes; this is supported by preliminary X-ray measurements, and also by analogy with the CeNi(Ge$_{1-x}$Si$_x$)$_2$ series, in which the end members have the same structure.  A similar saturation of $T_N$ was also found in \cng\ by Ohashi et al \cite{Ohashi06,Ohashi07} in a large-volume cell with a liquid medium, so it is unlikely that poor pressure conditions are responsible for the observed behavior.

In conclusion, the evolution of magnetic order in \cng\ appears to contradict the Doniach model, being very different from that seen in compounds considered archetypical of the Ce-based HF family such as CeIn$_3$. It has been observed experimentally that bare QCPs are rarely if ever found in real systems, and tend to be obscured by some other exotic phase.  It may be that \cng\ is a further example of this phenomenon.



\end{document}